\documentclass[a4paper,11pt]{article}
\usepackage{pos}
\usepackage{setspace}
\usepackage{wrapfig}

\usepackage{cleveref}
\Crefformat{equation}{Eq.~(#2#1#3)}
\Crefmultiformat{equation}{Eqs.~(#2#1#3)}{ and~(#2#1#3)}{, (#2#1#3)}{ and~(#2#1#3)}
\Crefformat{figure}{Fig.~#2#1#3}
\Crefformat{section}{Sec.~#2#1#3}

\allowdisplaybreaks

\def\pdotq{p \cdot q}

\def\latmom{\left( \frac{2\pi}{L} \right)}

\newcommand{\bv}[1]{{\bf{#1}}}
\newcommand{\bs}[1]{{\boldsymbol{#1}}}
\newcommand{\qb}{\bv{q}}
\newcommand{\pb}{\bv{p}}
\newcommand{\bl}{\bs{\lambda}}

\title{The parity-odd structure function of nucleon from the Compton amplitude}

\author*[a]{K.~U.~Can}
\author[b]{R.~Horsley}
\author[c]{Y.~Nakamura}
\author[d]{P.~E.~L.~Rakow}
\author[e]{G.~Schierholz}
\author[f]{H.~St\"{u}ben}
\author[a]{R.~D.~Young}
\author[a]{J.~M.~Zanotti}

\affiliation[a]{CSSM, Department of Physics, The University of Adelaide,
Adelaide SA 5005, Australia.}
\affiliation[b]{School of Physics and Astronomy, University of Edinburgh, Edinburgh EH9 3JZ, UK.}
\affiliation[c]{RIKEN Center for Computational Science, Kobe, Hyogo 650-0047, Japan.}
\affiliation[d]{Theoretical Physics Division, Department of Mathematical Sciences, University of Liverpool, Liverpool L69 3BX, United Kingdom.}
\affiliation[e]{Deutsches Elektronen-Synchrotron DESY, Notkestr. 85, 22607 Hamburg, Germany.}
\affiliation[f]{Regionales Rechenzentrum, Universit\"{a}t Hamburg, 20146 Hamburg, Germany.}

\emailAdd{kadirutku.can@adelaide.edu.au}

\abstract{The dominant contribution to the theoretical uncertainty in the extracted weak parameters of the Standard Model comes from the hadronic uncertainties in the electroweak boxes, i.e. $\gamma-W^\pm/Z$ exchange diagrams. A dispersive analysis relates the box diagrams to the parity-odd structure function, $F_3$, for which the experimental data either do not exist or belong to a separate isospin channel. Therefore a first-principles calculation of $F_3$ is highly desirable. 

In this contribution, we report on the QCDSF/UKQCD Collaboration's progress in calculating the moments of the $F_3^{\gamma Z}$ structure function from the forward Compton amplitude at the SU(3) symmetric point.}

\FullConference{The 40th International Symposium on Lattice Field Theory (Lattice 2023)\\
July 31st - August 4th, 2023\\
Fermi National Accelerator Laboratory\\}

\begin{document}
\maketitle

\section{Introduction}
Deep-inelastic neutrino-nucleon scattering, involving the exchange of a photon and a heavy $W/Z$ gauge boson, i.e. the electroweak box diagram, contains important information on radiative corrections of weak decays and parity violating lepton-nucleon interactions. Radiative corrections are one of the main sources of theoretical uncertainty in the determination of the CKM matrix elements and the weak mixing angle at low scales~\cite{Hardy:2014qxa,Kumar:2013yoa}. Precise parity-odd structure functions are of vital importance to the neutrino oscillation experimental program~\cite{Kronfeld:2019nfb}, particularly at large Bjorken-$x$ and at the highest energies of the DUNE~\cite{DUNE:2015lol} beam. Furthermore, weak currents, unlike the electromagnetic current, couple nontrivially to spin and flavour~\cite{Forte:2005pv,Ball:2009mk}, thus providing useful constraints on the spin and flavour decomposition of parton distribution functions (PDFs)~\cite{Ball:2009mk}. 

Lattice QCD is well-positioned to provide first-principles determinations of the structure functions. We have been pursuing the calculations of the moments of structure functions from the Compton amplitude calculated by a Feynman-Hellmann approach initiated in~\cite{PhysRevLett.118.242001} and applied to unpolarised~\cite{PhysRevD.102.114505,QCDSFUKQCDCSSM:2022ncb} and polarised~\cite{Can:2022chd} parity-conserving nucleon structure functions. In this contribution, we expand our calculations to include the unpolarised parity-odd structure function. The lowest moment of the parity-odd structure function is of particular importance since it can be related to the electroweak box diagram contribution via a dispersion relation,
\begin{equation} \label{eq:disp_gW}
  \square_{VA}^{\gamma W} = \frac{3 \alpha_{EM}}{2\pi} \int_0^\infty \frac{dQ^2}{Q^2} \frac{M_W^2}{M_W^2+Q^2} M_1^{(3)}(Q^2),
\end{equation}
where $\alpha_{EM}$ is the fine-structure constant, $M_W$ is the mass of the $W$ boson, and $M_1^{(3)}$ is the first Nachtmann moment of the flavour non-diagonal structure function $F_3^{\gamma W}$. A recent lattice QCD calculation of the $\gamma W$-box contribution to superallowed nuclear and neutron beta decays is presented in~\cite{Ma:2023kfr}. A dispersion relation analogous to \Cref{eq:disp_gW} can be written for the $\gamma Z$ box diagram. Isospin symmetry provides the relation, $F_3^{\gamma W} \equiv (F_3^{\gamma Z,p} - F_3^{\gamma Z,p})/4$, between the flavour non-diagonal and flavour diagonal structure functions, where the calculation of the $Q^2$ dependence of the lowest moment of $F_3^{\gamma Z}$ using a Feynman-Hellmann approach is the focus of this contribution. 

Apart from its relation to the neutrino-nucleon deep-inelastic scattering phenomena discussed above, the first moment of the $\gamma Z$ interference structure function provides a significant test of the Gross–Llewellyn-Smith (GLS) sum rule. In the quark-parton model the first moment of $F_3^{\gamma Z}$ is equal to the number of valence quarks in the nucleon, it is a non-singlet quantity, and it has a vanishing anomalous dimension in QCD~\cite{Retey:2000nq}, making the GLS sum rule a very clean quantity to study. The QCD corrections for the leading twist contribution of the GLS sum rule has been computed to NNLO~\cite{Larin:1991tj,Retey:2000nq} and the power corrections have been addressed in model calculations~\cite{Braun:1986ty,Ross:1993gb,Dasgupta1996}, although with widely varying results. A first-principles calculation would therefore provide insights to the interplay between the perturbative and nonperturbative regimes.

The rest of this paper is organised as follows: In \Cref{sec:ca} we discuss the decomposition of the Compton amplitude and the kinematics to isolate the parity-odd Compton structure function $\mathcal{F}_3$. A brief summary of the Feynman-Hellmann approach employed to calculate the Compton amplitude is given in \Cref{sec:fh}, followed by the preliminary results in \Cref{sec:res}. Our concluding remarks are given in \Cref{sec:sum}.

\section{Parity-odd Compton amplitude} \label{sec:ca}
The starting point of the calculation is the time-ordered product of vector and axial vector currents, $J_\mu$ and $J_\nu^A$, sandwiched between nucleon states, forming the hadronic tensor,
\begin{equation} \label{eq:htensor}
  T_{\mu\nu}(p,q) = i \rho_{s s^\prime} \int d^4z e^{i q \cdot z} \langle p,s^\prime | \mathcal{T}\{J_\mu(z) J_\nu^A(0)\} | p,s \rangle,
\end{equation} 
where $p$ ($q$) is the nucleon (current) momentum, and $\rho_{s s^\prime}$ is the spin density matrix. The Lorentz decomposition of the hadronic tensor for the spin-averaged case involves six structure functions~\cite{Ji:1993ey,Thomas:2001kw},
\begin{align}
  T_{\mu\nu}(p,q) =& -g_{\mu\nu} \mathcal{F}_1(\omega,Q^2) 
  + \frac{p_\mu p_\nu}{\pdotq} \mathcal{F}_2(\omega,Q^2) 
  + i \, \varepsilon_{\mu\nu}\!{}^{\alpha\beta} \frac{p_\alpha q_\beta}{2 \pdotq}\mathcal{F}_3(\omega,Q^2) \nonumber \\
  &+ \frac{q_\mu q_\nu}{\pdotq} \mathcal{F}_4(\omega,Q^2) 
  + \frac{p_{{}_{\{}\mu} q_{\nu_{\}}}}{\pdotq} \mathcal{F}_5(\omega,Q^2) 
  + \frac{p_{{}_{[}\mu} q_{\nu_{]}}}{\pdotq} \mathcal{F}_6(\omega,Q^2),
\end{align}
where $\varepsilon^{0123} = 1$, $\omega = 2 \pdotq / Q^2$, and $p_{{}_{\{}\mu} q_{\nu_{\}}} = (p_\mu q_\nu + p_\nu q_\mu)/2$ and $p_{{}_{[}\mu} q_{\nu_{]}} = (p_\mu q_\nu - p_\nu q_\mu)/2$ denote the symmetrisation and antisymmetrisation of the indices respectively. Here, $\mathcal{F}_{1,2}$ are the familiar unpolarised Compton structure functions. Since the axial charge is not conserved and the parity is violated, there are additional structure functions. $\mathcal{F}_{4,5}$ are not related to $\mathcal{F}_{1,2}$ by gauge invariance any more so they survive, and $\mathcal{F}_{3,6}$ are allowed because the parity is violated, although $\mathcal{F}_{6}=0$ due to the time reversal invariance of strong interactions.

We are interested in the $\mathcal{F}_3$ amplitude, which is the only surviving parity-violating part. Choosing the off-diagonal components of the tensor, e.g. $\mu \ne \nu$, and $p_\mu = q_\mu = 0$, it is straightforward to isolate $\mathcal{F}_3$,
\begin{equation} \label{eq:f3}
  T_{\mu\nu}(p,q) = i \, \varepsilon_{\mu\nu}\!{}^{\alpha\beta} \frac{p_\alpha q_\beta}{2 \pdotq}\mathcal{F}_3(\omega,Q^2),
\end{equation}
which is connected to the unpolarised parity-violating structure function $F_3$ through the dispersion relation,
\begin{equation} \label{eq:disp}
  \mathcal{F}_3(\omega, Q^2) = 4 \omega \int_0^1 dx \frac{F_3(x,Q^2)}{1-x^2 \omega^2}.
\end{equation}
Upon expanding the geometric series we arrive at,
\begin{equation} \label{eq:ca_expand}
  \frac{\mathcal{F}_3(\omega, Q^2)}{\omega} = 4 \sum_{n=1,2,\dots} \omega^{2n-2} \, M_{2n-1}^{(3)}(Q^2),
\end{equation}
with the odd Mellin moments of $F_3$ defined as,
\begin{equation} \label{eq:moments}
  M_{2n-1}^{(3)}(Q^2) = \int_0^1 dx \, x^{2n-2} \, F_3(x,Q^2), \quad \text{for } n=1,2,3, \dots
\end{equation}
We note that the lowest Mellin moment, $M_1^{(3)}(Q^2)$, is directly accessible at $\omega=0$, i.e. for a nucleon at rest, thus can be extracted from the Compton amplitude without a polynomial fit. 

\section{Feynman-Hellmann approach} \label{sec:fh}
Now the task is to calculate the Compton amplitude. An analysis of the Compton amplitude requires the evaluation of lattice four-point correlation functions. However, this is not an easy task given the rapid deterioration of the signal for large time separations and the contamination due to excited states. The application of the Feynman-Hellmann theorem reduces the problem to a simpler analysis of two-point correlation functions using the established techniques of spectroscopy. Our implementation of the second order Feynman-Hellmann method is presented in detail in~\cite{PhysRevD.102.114505}. 

To compute the hadronic tensor, \Cref{eq:htensor}, that involves the product of a vector and axial vector currents, the Feynman-Hellmann technique needs to be generalised to mixed currents which has been done in the context of generalised parton distribution calculations before \cite{Alec:2021lkf}, albeit for electromagnetic currents. It is also clear from \Cref{eq:f3} that we need the antisymmetric part of the tensor. In this case, we introduce two spatially oscillating background fields to the action
\begin{equation}\label{eq:fh_action}
  S(\bl) = S_0 
  + \lambda_1 \int d^4z \operatorname{cos}(q \cdot z) \mathcal{J}_\mu(z)
  + \lambda_2 \int d^4y \operatorname{sin}(q \cdot y) \mathcal{J}^A_\nu(y),
\end{equation}
where $S_0$ is the unperturbed action, $\lambda_1$, $\lambda_2$ are the strengths of the couplings between the quarks and the external fields, and $\bl \equiv (\lambda_1,\lambda_2)$; $\mathcal{J}_\mu(z) = Z_V \bar q(z) \gamma_\mu q(z)$ and $\mathcal{J}_\nu^A(y) = Z_A \bar q(y) \gamma_5 \gamma_\nu q(y)$ are the renormalised vector and axial vector currents coupling to the quarks along the $\mu$ and $\nu$ directions, $q=(\qb,0)$ is the external momentum inserted by the currents and $Z_{V,A}$ are the renormalisation constants for the local vector and axial vector currents. We note that a multiplicative renormalisation is the only renormalisation that is needed in the Feynman-Hellmann approach where $Z_{V,A}$ are determined before~\cite{Constantinou:2014fka}. Following the derivation presented in~\cite{PhysRevD.102.114505}, we arrive at the Feynman-Hellmann relation between the second-order energy shift and the Compton amplitude, 
\begin{equation} \label{eq:secondorder_fh}
    \left. \frac{\partial^2 E_{N_\lambda}(\pb, \qb)}{\partial \lambda_1 \partial \lambda_2} \right|_{\bl=0} 
    = i \frac{T_{\mu\nu}(p,q) - T_{\mu\nu}(p,-q)}{2 E_{N}(\pb)} 
    = \frac{i T_{\mu\nu}(p,q)}{E_{N}(\pb)},
\end{equation}
where $T$ is the Compton amplitude defined in \Cref{eq:htensor}, and $E_{N_\lambda}(\pb, \qb)$ is the nucleon energy at momentum $\pb$ in the presence of a background field of strength $\bl$. For an independent derivation, based on an expansion of the Lagrangian in terms of a periodic external source~\cite{Agadjanov:2016cjc}, see~\cite{Seng:2019plg}.

\section{Preliminary results} \label{sec:res}
In order to isolate the parity-odd Compton structure function $\mathcal{F}_3$, we choose the kinematics $\mu=1$, $\nu=3$, $\alpha=2$, $\beta=0$, $\pb_1=\qb_1=0$ and $\qb_2 \ne 0$, and noting that $q_0 = 0$ in the FH approach, we rewrite \Cref{eq:f3} as
\begin{equation} \label{eq:f3ip}
    \frac{\mathcal{F}_3(\omega,Q^2)}{\omega} = \frac{Q^2}{\qb_2} \left. \frac{\partial^2 E_{N_{\lambda}}(\pb)}{\partial \lambda_1 \partial \lambda_2} \right|_{\bl=0},
\end{equation}  
with the use of $\omega = 2 \pdotq / Q^2$ and \Cref{eq:secondorder_fh}. Here we remind the reader that in practice, we determine the RHS of \Cref{eq:f3ip} so that we do not explicitly divide the LHS by $\omega$, hence no singularity at $\omega=0$ occurs. 

To calculate the second-order energy shift, we construct the ratio,
\begin{align} \label{eq:ratio}
  \mathcal{R}_{\lambda}(\pb,\qb,t) &\equiv 
  \frac{
  G^{(2)}_{+\lambda_1,+\lambda_2}(\pb,\qb,t) \, 
  G^{(2)}_{-\lambda_1,-\lambda_2}(\pb,\qb,t)
  }
  {
  G^{(2)}_{+\lambda_1,-\lambda_2}(\pb,\qb,t) \, 
  G^{(2)}_{-\lambda_1,+\lambda_2}(\pb,\qb,t)
  } 
  \xrightarrow{t \gg 0} A_\lambda(\pb) e^{-4\Delta E^{oo}_{N_\lambda}(\pb,\qb) \, t},
\end{align}
which extracts the energy shift $\Delta E^{oo}_{N_{\lambda}}(\pb,\qb)$, where $A_\lambda(\pb)$ is the overlap factor irrelevant for the rest of the discussion. The ratio in \Cref{eq:ratio} isolates the energy shift only at even orders of $\bl$, e.g. $\mathcal{O}(\lambda_1^n \lambda_2^m)$ with $n+m$ even and $n,m \ge 0$. Here, $G^{(2)}_{\lambda_1, \lambda_2}(\pb,\qb,t) \xrightarrow{t \gg 0} A^{\prime}_\lambda(\pb) e^{-E_{N_{\lambda}}(\pb,\qb) t}$, is the perturbed two-point function with $|\lambda_1| = |\lambda_2| = |\lambda|$, where $A_\lambda(\pb)$ is the overlap factor and $E_{N_{\lambda}}(\pb,\qb)$ is the perturbed energy of the ground state. The relation between $\Delta E^{oo}_{N_{\lambda}}(\pb,\qb)$ and the Compton amplitude is rendered visible when considering the perturbed energy of the nucleon expanded as a Taylor series in the limit $\bs{\lambda} \to 0$,
\begin{equation}
  E_{N_{\lambda}}(\pb,\qb) = E_N(\pb) 
    + \Delta E^{eo}_{N_{\lambda}}(\pb,\qb) 
    + \Delta E^{oe}_{N_{\lambda}}(\pb,\qb) 
    + \Delta E^{ee}_{N_{\lambda}}(\pb,\qb) 
    + \Delta E^{oo}_{N_{\lambda}}(\pb,\qb),
\end{equation}
where the superscript $e$ ($o$) denote the terms even (odd) in $\lambda_{1,2}$. The term of interest is,
\begin{equation} \label{eq:enshift_oo}
    \Delta E^{oo}_{N_{\lambda}}(\pb,\qb) = \lambda_1 \lambda_2 \left. \frac{\partial^2 E_{N_{\lambda}}(\pb,\qb)}{\partial \lambda_1 \partial \lambda_2} \right|_{\bl=0} + \mathcal{O}(\lambda_1 \lambda_2^3) + \mathcal{O}(\lambda_1^3 \lambda_2),
\end{equation}
which holds the energy shift associated with the interference of the currents $\mathcal{J}_\mu$ and $\mathcal{J}_\nu^{A}$ appearing on the LHS of \Cref{eq:secondorder_fh}. 

Our calculations are performed on QCDSF/UKQCD-generated $2+1$-flavour gauge configurations. One ensemble is used with volume $V=48^3 \times 96$, and coupling $\beta=5.65$ corresponding to the lattice spacing $a=0.068 \, {\rm fm}$. Quark masses are tuned to the $SU(3)$ symmetric point where the masses of all three quark flavours are set to approximately the physical flavour-singlet mass, $\overline{m} = (2 m_s + m_l)/3$~\cite{Bietenholz:2010jr,Bietenholz:2011qq}, yielding $m_\pi \approx 420 \, {\rm MeV}$. As these are preliminary calculations we restrict our calculations to a fixed photon virtuality of $Q^2 \sim 5 \, {\rm GeV}^2$ which is obtainable via two different choices of $\qb = [(0,3,5), \, (0,5,3)] \latmom$. We perform a low statistics run of approximately $500$ and $250$ measurements of the ratio (\Cref{eq:ratio}) for the two $\qb$ vectors, respectively. 
Since $F_3$ is a non-singlet quantity we only calculate the connected diagrams. 

\begin{figure}[t]
    \centering
    \includegraphics[width=.7\textwidth]{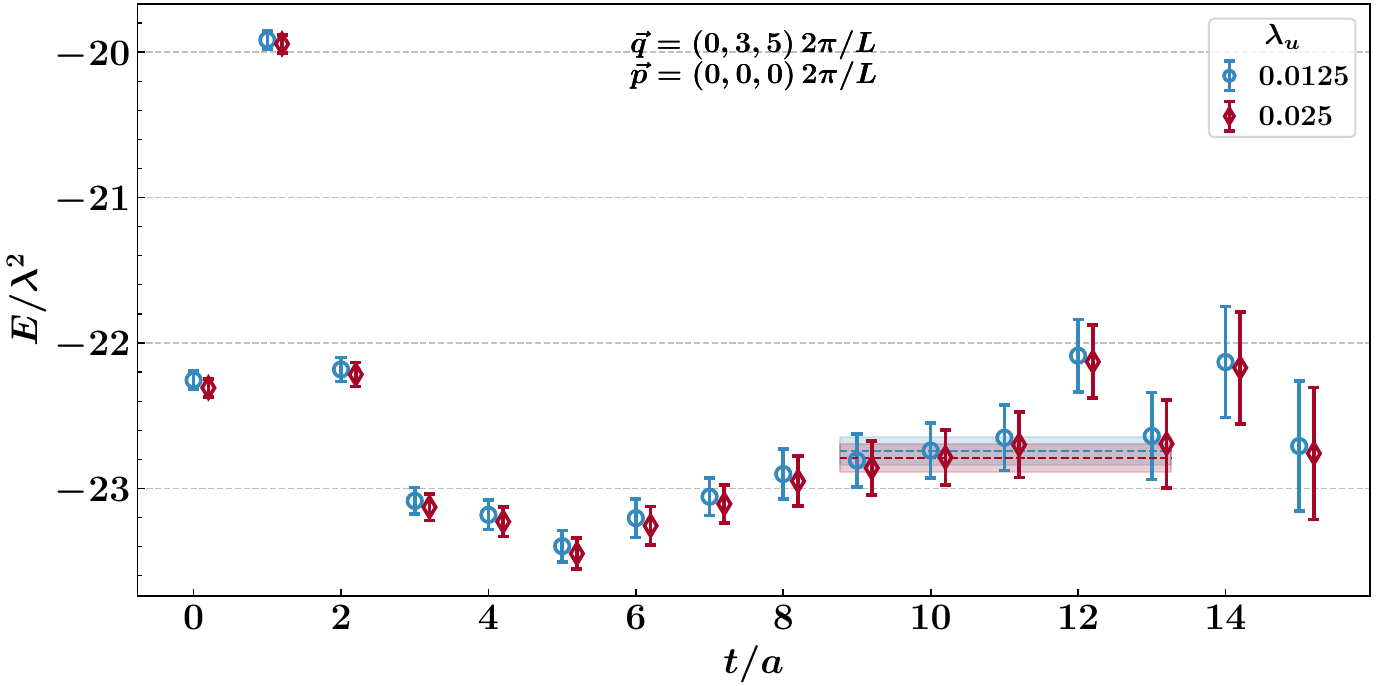}
    \caption{\label{fig:effmass}Effective mass plot of the ratio (\Cref{eq:ratio}). Fit windows, along with the extracted energy shifts with their $1\sigma$ uncertainty, are shown by the shaded bands. We are showing the results for the $uu$ contribution, for $(\bv{p},\bv{q}) = ((0,0,0),(0,3,5)) \, \latmom$ corresponding to $\omega = 0.0$ at $Q^2 \sim 5 \, {\rm GeV}^2$. 
    }
\end{figure}

\begin{figure}[t]
    \centering
    \includegraphics[width=.7\textwidth]{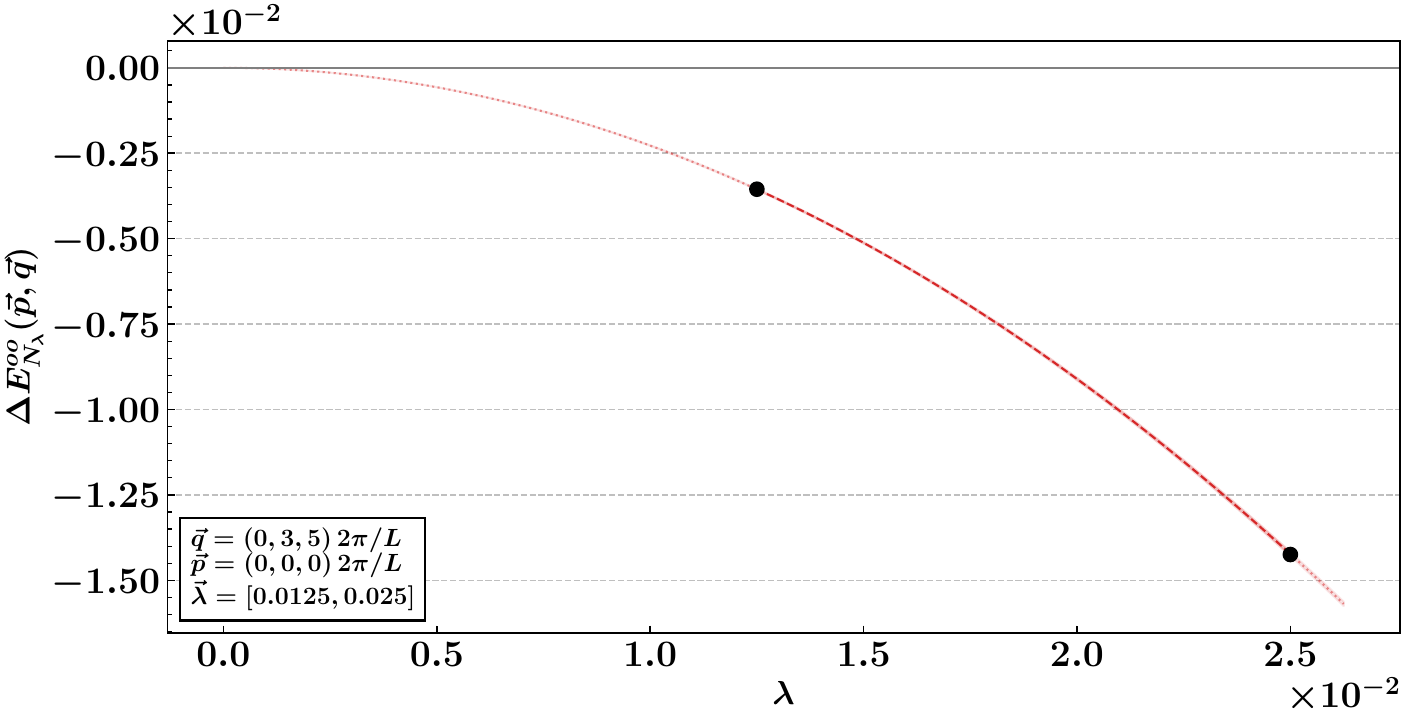}
    \caption{\label{fig:lamfit}$\lambda$ dependence of the extracted energy shift for the same kinematics given in \Cref{fig:effmass}. The uncertainty of the points are smaller than the symbols.
    }
\end{figure}

All the energy shifts, $\Delta E^{oo}_{N_{\lambda}}(\pb,\qb)$, are extracted by a fit to the ratio given in \Cref{eq:ratio}. We show a representative effective mass plot for the ratio for a nucleon at rest, $\pb = (0,0,0) \latmom$, at $\qb = (0,3,5) \latmom$ in \Cref{fig:effmass}. The fit windows are determined by a covariance-matrix based $\chi^2$ analysis where the chosen ranges are required to have $\chi^2_{dof} \sim 1.0$. To map the $\bs{\lambda}$ dependence of the energy shift we calculate the ratio for $|\bs{\lambda}| = [0.0125, 0.025]$ for each combination of $\pb$ and $\qb$. Then, we perform a polynomial fit using \Cref{eq:enshift_oo} to determine $\left. \frac{\partial^2 E_{N_{\lambda}}(\pb)}{\partial \lambda_1 \partial \lambda_2} \right|_{\bl=0}$. We do not find a statistically significant contamination due to neglected higher-order terms in \Cref{eq:enshift_oo}. Our resulting fit is shown in \Cref{fig:lamfit}.

\begin{figure}[h]
    \centering
    \includegraphics[width=\textwidth]{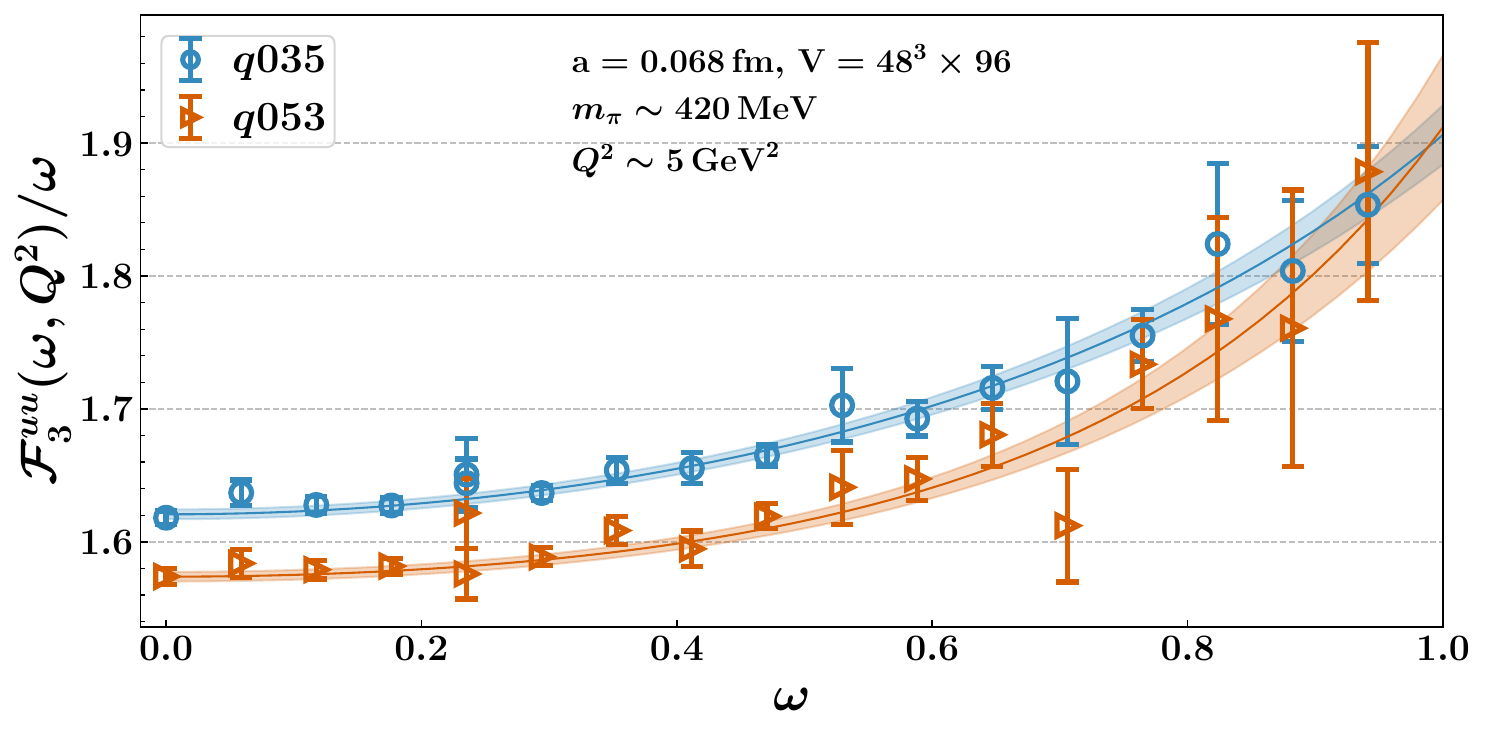}
    \caption{\label{fig:f3}$\omega$ dependence of $\mathcal{F}_3(\omega, Q^2)$ at each simulated value of $\bv{q}$ for the $uu$ contribution. 
    }
\end{figure}

We map the $\omega$ dependence of $\mathcal{F}_3(\omega, Q^2)$ at each $\bv{q}$ by varying the nucleon momentum $\bv{p}$ and following the analysis outlined above. At our chosen kinematics we access 18 evenly distributed $\omega$ values providing a good coverage of the range $0 \leq \omega \lesssim 1$. The resulting $\mathcal{F}_3(\omega, Q^2)$ for $\qb = [(0,3,5), \, (0,5,3)] \latmom$ are shown in \Cref{fig:f3}. Even though we have modest statistics, we determine the $\omega=0$ point, i.e. the lowest moment, with sub-percent precision. The shaded bands are from a polynomial fit (\Cref{eq:ca_expand}), whose details can be found in~\cite{PhysRevD.102.114505,QCDSFUKQCDCSSM:2022ncb}, performed to extract the first few moments of the $F_3(x,Q^2)$ structure function.  At this stage however, we refrain from quoting any results for the moments before the apparent lattice artefacts are addressed. We are working towards controlling the systemtic effects.

\section{Summary and outlook} \label{sec:sum}
In this contribution we have reported on our preliminary calculations of the parity-violating Compton amplitude using an extension of the Feynman-Hellmann approach. Our exploratory calculations have been performed on a single $2+1$ flavour, $48^3 \times 96$ gauge ensemble with $\beta=5.65$, corresponding to a lattice spacing $a=0.068$, with the light quark masses tuned to approximately the physical flavour-singlet mass yielding $m_\pi \approx 420 \, {\rm MeV}$. We have mapped the $\omega$-dependence of the parity-odd Compton structure function $\mathcal{F}_3$ for a fixed $Q^2$ but with two different $\qb$ vector geometries. Our results reach a good statistical precision and reveal the effects of the lattice artefacts. We are working towards controlling these systematic effects. Once the systematic uncertainties are addressed, our results would provide valuable input for estimating the electroweak box-diagram contributions to the superallowed nuclear and neutron beta decays and testing the Gross–Llewellyn-Smith sum rule.

{\small
\acknowledgments
The numerical configuration generation (using the BQCD lattice QCD program~\cite{Haar:2017ubh})) and data analysis (using the Chroma software library~\cite{Edwards:2004sx}) was carried out on the DiRAC Blue Gene Q and Extreme Scaling (EPCC, Edinburgh, UK) and Data Intensive (Cambridge, UK) services, the GCS supercomputers JUQUEEN and JUWELS (NIC, Jülich, Germany) and resources provided by HLRN (The North-German Supercomputer Alliance), the NCI National Facility in Canberra, Australia (supported by the Australian Commonwealth Government) and the Phoenix HPC service (University of Adelaide). RH is supported by STFC through grant ST/P000630/1. PELR is supported in part by the STFC under contract ST/G00062X/1. KUC, RDY and JMZ are supported by the Australian Research Council grants DP190100297 and DP220103098. For the purpose of open access, the authors have applied a Creative Commons Attribution (CC BY) licence to any Author Accepted Manuscript version arising from this submission.
}
\begingroup
    \small
    \renewcommand{\baselinestretch}{1}
    \setlength{\bibsep}{0pt}
    \setstretch{1}

    \providecommand{\href}[2]{#2}\begingroup\raggedright\endgroup

\endgroup

\end{document}